\documentclass[12pt, letterpaper]{article}
\usepackage{geometry}
\usepackage{amsmath}
\usepackage{graphicx}
\usepackage{float}
\usepackage{siunitx}
\usepackage{subcaption}
\usepackage{amssymb}
\usepackage[utf8]{inputenc}
\usepackage[USenglish]{babel}
\newgeometry{vmargin={18mm}, hmargin={18mm,18mm}}
\usepackage{mathtools}
\DeclarePairedDelimiter\bra{\langle}{\rvert}
\DeclarePairedDelimiter\ket{\lvert}{\rangle}
\DeclarePairedDelimiterX\braket[2]{\langle}{\rangle}{#1\,\delimsize\vert\,\mathopen{}#2}

\providecommand{\keywords}[1]
{
  \small	
  \textbf{\textit{Keywords---}} #1
}
\begin{document}
\title{Quasi-stationary evolution of cubic-quintic NLSE drop-like solitons in DNA-protein systems}
\author{O. Pav\'on-Torres$^{1}\footnote{omar.pavon@cinvestav.mx}$, J. R. Collantes-Collantes$^{2}\footnote{juan.collantes@utp.ac.pa}$, M.A. Agüero-Granados$^{3}\footnote{maaguerog@uaemex.mx}$\\}
\date{%
\small{$^{1}$ Physics Department, Cinvestav, POB 14-740, 07000 Mexico City, Mexico\\
\small {$^{2}$ Facultad de Ciencias y Tecnolog\'ia, Universidad Tecnol\'ogica de Panam\'a, Apdo. 0819-07289, \\Panam\'a, Rep\'ublica de Panam\'a\\
\small{$^{3}$ Facultad de Ciencias, Instituto literario 100, Toluca, 5000, Mexico, Mexico}}
}}
\maketitle
\begin{abstract}
Nonlinear molecular excitations in DNA have traditionally been studied within the framework of the nonlinear Schrödinger equation (NLSE).  An alternative approach is based on the plane-base rotator model and $SU(2)/U(1)$ generalized spin coherent states, which lead to a cubic-quintic NLSE. Higher-order nonlinearities are particularly useful for modelling complex interactions, such as those in DNA-protein systems, where multiple competing forces play a significant role. Additionally, the surrounding viscous medium introduces dissipative forces that influence the propagation of molecular excitations, resulting in energy dissipation and damping effects. These damping effects are modelled using the quasi-stationary method, which describes the system's near-equilibrium behaviour. In this work, we analyse the evolution of nonlinear molecular excitations in DNA-protein systems, accounting for damping effects, and discuss potential applications to the transcription process. 
\end{abstract}


\section{Introduction}
The local physical properties of genomic DNA, including base pair deformations, impurities, and domain walls, are crucial in regulating stability, gene expression, sequence recognition, and protein binding \cite{00, DNA2, 0, 1, 2, 3, 4, 5}. For example, DNA undergoes sequence-dependent conformational changes, such as kinking and intercalation, in response to protein binding \cite{l0, l1}. This is particularly important in processes like chromosome condensation and segregation, which rely on DNA-binding proteins that facilitate chromosome organization through DNA bridging and bending \cite{l2}. Thus, the sequence-specific recognition of DNA by proteins and other ligands requires a more nuanced understanding of DNA's flexibility—an aspect that traditional uniform elastic models fail to fully capture. Therefore, accurately describing, analyzing, and representing these conformational changes demand the development of new ones \cite{DNAm3, DNA1, DNA2p, DNA3, DNA4, DNA5, DNA6, DNA7, DNAA0, DNAA1}. These conformational changes can be depicted as long-lived and large amplitude localized nonlinear excitations. More broadly, it is well-established that such excitations play a key role in a variety of biological processes involving DNA, including transcription, gene expression, and the recognition of promoter sites, among others \cite{l3}.
Experimentally, ensemble techniques such as surface plasmon resonance, white light interferometry, and electrophoretic mobility shift assays are commonly used to measure protein-DNA binding. However, these methods do not provide insight into how proteins induce conformational changes in the DNA \cite{song}. To investigate these structural changes, a more controlled environment is required, and DNA-designed crystals have emerged as a promising tool for this purpose \cite{cry5, cry6, cry1, cry2, cry3, cry4}. This innovative approach enables a detailed analysis of DNA’s structural and physical properties, along with its interactions and behaviour, thereby advancing our understanding of the macromolecule. As this field continues to grow, it is expected to yield new insights into DNA-protein dynamics soon \cite{cry7}. 

Despite significant efforts, studying DNA conformational changes resulting from interactions with other molecules remains a long-standing challenge. However, this can be addressed through the implementation of mathematical models. In this context, the double-stranded model of DNA, introduced by Takeno and Homma \cite{take1, take2} and commonly referred to as the plane base-rotator model, has become widely accepted. Using a generalized version of this model: M. Daniel and his collaborators devoted a series of studies to the internal dynamics of inhomogeneous double-stranded DNA molecular system, which dynamics have been governed by the sine-Gordon equation and the NLSE \cite{vasum, danvas, gre1, dani, vasum2}; Ag\"uero et al, using the generalized coherent states approach, in Perelomov sense, studied the DNA-protein interaction obtaining several classical and non-classical soliton solutions that describe the open states of base pairs \cite{max1, max2, max3}; Saha and Kofan\'e employed the plane base-rotator model to study numerically the effects of inhomogeneities on the double stranded DNA interacting with RNAp enzyme \cite{kofa1, kofa2};  Saccomandi studied the inhomogeneous stacking on the double strand using the Peyrard-Bishop model and Takeno-Homma model \cite{saco}. 

In most of the studies mentioned above, the internal dynamics of base pairs, influenced by protein interactions, were described using NLSE with different nonlinearities. This kind of equation have widely extended in many branches of mathematical physics in order to explain nonlinear phenomena like water waves or the propagation of light pulses in fibres for example  \cite{triki, zayed, ref1, ref3, craik, kurk, ref2, ref4, gou, gedal}. 
A very particular case of interest is the NLSE with cubic-quintic nonlinearities commonly known as the cubic-quintic NLSE, which appears in many branches of physics ranging from Bose-Einstein condensates to the description of optical pulse propagations in dielectric media of non-Kerr type, including plasma physics, nuclear hydrodynamics, generalized elastic solids, among others \cite{kart, gam, bongs, zhou, irm, kum, ref5, sref1, sref2, sref3, sref4, sref5, sref5, sref6, sref7, sreff0, sreff1, sreff2, sref8, sref9, sref10, sref11, sref12, sref13, sref14, sref15}. 

Another important contribution to a more accurate description of the DNA-protein system and its associated nonlinear excitations is the inclusion of a viscous medium as a damping term. The addition of this term mathematically leads to non-integrability of the resulting equation. Thus, exact $N$- soliton solutions cannot be obtained, and a perturbative analysis becomes necessary \cite{vasum}. In the case of the NLSE with higher-order nonlinearities, the non-integrability is less obvious, since Lax pairs cannot be directly constructed, and thus no $N$-soliton solutions can be derived. Such equations arise when the generalized coherent states approach is employed, leading to a cubic-quintic NLSE. The aim of the present work is to investigate the evolution of nonlinear excitations in the DNA-protein molecular system embedded in a viscous medium. This analysis is performed using a multi-scale perturbation approach, which allows for the identification of localized perturbations around the soliton. Through this method, we observed how the soliton parameters are modified by the perturbative term. A similar study was conducted in the context of optical fibres \cite{serk0}, but the approach can also be extended to biological systems, such as lipid membranes in the framework of a Boussinesq-like equation \cite{omm1}. Motivated by these previous works, we present this analysis, which is organized as follows: In Section 2, we provide a brief overview of the plane base-rotator model, where the dynamics of the DNA base pairs with a binding protein are described by the equations of motion derived from the generalized Hamiltonian. After some simplifications, these equations are reduced to the NLSE with cubic-quintic nonlinearities, facilitated by the use of $SU(2)/U(1)$ generalized coherent states. Section 3 focuses on the interaction between the DNA and the surrounding medium. As is well known, the viscous medium introduces dissipative forces that affect the propagation of molecular excitations, leading to energy dissipation and damping effects. To model the steady-state or near-equilibrium behavior of the system, we adopt the quasi-stationary approximation, in which the system evolves slowly relative to the dissipation timescale. In Section 4, we apply these analytical results to the transcription process. Finally, Section 5 concludes the paper with a summary and discussion of our findings.
 
\section{Nonlinear molecular excitations in DNA-protein system}
\subsection{Plane-base rotator model of a DNA-protein system}

Let us consider the B-form of DNA, where the sugar-phosphate backbone is represented by two ribbons, $S$ and $S'$, that spiral around each other along the $z$-axis, as illustrated in Fig. 1a \cite{take1, take2}. The coordinates $P_{n}$ and $P_{n}'$ represent the points where the $n$-th and $n'$-th bases are attached to their respective strands. These points are given by $(r \cos(n\varphi_{0}),r \sin(n\varphi_{0}), z_{n})$ and $(r \cos(n\varphi_{0}+\pi),r \sin(n\varphi_{0}+\pi), z_{n}')$, where $\varphi_{0}=2\pi/p$. Here, $r$ is the radius of the circle and $p=10$ is the number of bases per turn in $S$ and $S'$, respectively, as shown in Fig. 1b. 
Additionally, $(\theta_{n}, \varphi_{n})$ and $(\theta_{n}', \varphi_{n}')$ represent the angles of rotation of the $n$-th and $n'$-th base pairs around the points $P_{n}$ and $P_{n}'$, correspondingly.
\begin{figure}
\includegraphics[width=1.0\linewidth]{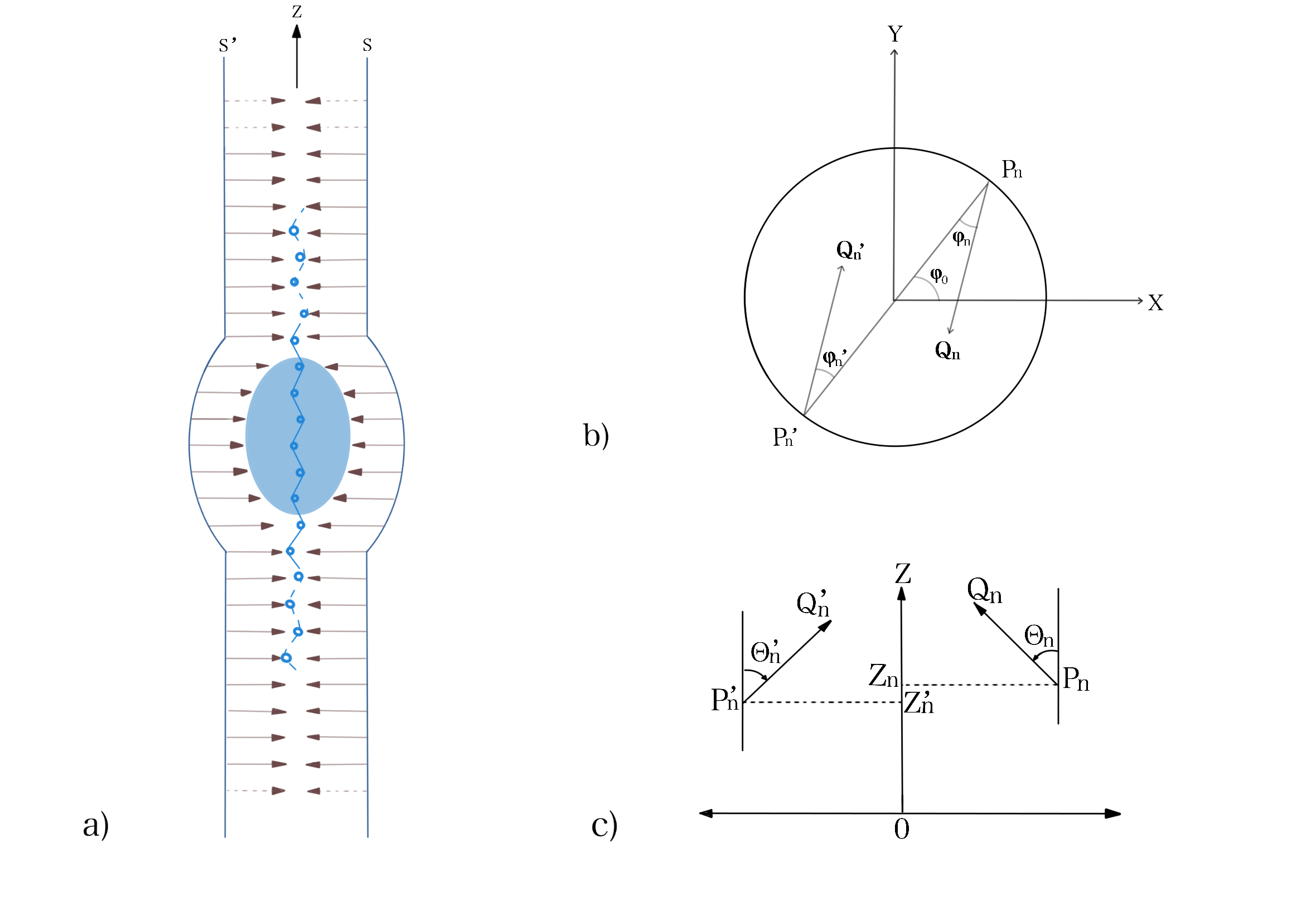}
\centering
\caption{a) Schematic representation of the DNA-protein system: DNA is modelled as two coupled linear molecular chains, while the protein is represented as a single linear chain interacting with DNA through a linear coupling. The shaded ellipse indicates the region where the protein molecule binds to a specific segment of the DNA. b) Projection of $n$-th and $n'$-th base pairs onto the $xy$-plane and c) Projection of $n$-th and $n'$-th base pairs onto the $xz$ plane.}
\label{fig:tak1}
\end{figure}
Below, we outline the general procedure for deriving the generalized Hamiltonian as an extension of the original plane-base rotator model (for further details, see \cite{gre1, max1}).
\begin{enumerate}
\item The hydrogen bond (or inter-strand) energy can be derived from the heuristic argument that it depends on the distance between the tip of the $n$-th base and its complementary $n'$-th base in the opposing strand, denoted as $\overline{Q_{n}Q_{n}'}$. For simplicity, we assume that the positions of the bases, $z_{n}=z_{n}'$, are identical, as shown in Fig. 1c.
\item The quasi-spin operators $\vec{S}_{n}=(S_{n} ^{x}, S_{n} ^{y}, S_{n}^{z})$, expressed in terms of the rotational angles $\theta_{n}$ and $\varphi_{n}$, with components
$$S_{n} ^{x}=\sin \theta_{n} \cos \varphi_{n}, \quad S_{n} ^{y}=\sin \theta_{n} \sin \varphi_{n} \quad \text{and} \quad S_{n} ^{z}=\cos \theta_{n} $$
and, its complementary $\vec{S}_{n}'$ are introduced. 
\item An analogy is drawn between double-stranded DNA and an anisotropic coupled spin chain model to derive the intra-strand (or stacking) energy.
\item The kinetic and potential energies of the DNA, along with the protein energy, are summed up, where $X_{n}$ stands for the $n$-th base displacement along the direction of the hydrogen bond, and $Y_{n}$ denotes the displacement for $n$-th peptide in the protein chain. The corresponding momenta are $p_{n}$ and $q_{n}$, respectively.  
\item Finally, the coupling between the DNA with protein is introduced. 
\end{enumerate} 
Taking into account all the aforementioned requirements, the following Hamiltonian is obtained:
\begin{eqnarray}
H &=& \displaystyle\sum_{n}[-J(\textbf{S}_{n}{\cdot}\textbf{S}_{n+1}+\textbf{S}_{n}'{\cdot}\textbf{S}_{n+1}')-(\mu-\alpha_{1}(X_{n+1}+X_{n-1}))(\textbf{S}_{n}{\cdot}\textbf{S}_{n+1})+\alpha_{2}(Y_{n-1}+Y_{n+1})(S_{z}S_{z}')\nonumber\\
&& +\: \frac{p_{n}^{2}}{2m_{1}}+k_{1}(X_{n}-X_{n+1})^{2}+\frac{q_{n}^{2}}{2m_{2}}+k_{2}(Y_{n}-Y_{n+1})^{2}], \label{darr4}
\end{eqnarray}
here, $J$ and $\mu$ are constants that represent the stacking interaction energy between the $n$-th base and its nearest neighbours in the two strands, and the hydrogen bond energy between the complementary bases, correspondingly. Additionally, $k_1$ and $k_2$ are elastic constants that characterize the behaviour of the hydrogen atom attached to the base and the amplitude of oscillations in the protein. The masses $m_1$ and $m_2$ correspond to the hydrogen atom attached to the base and the peptide in the protein, respectively. Finally, $\alpha_1$ and $\alpha_2$ are coupling coefficients that describe the interaction between thermal phonons, DNA and the protein.

\subsection{Weakly saturable approximation and cubic-quintic NLSE}

We can exploit the fact that the Hamiltonian (\ref{darr4}) is expressed in terms of quasi-spin operators, along with the generalized coherent states associated with the $SU(2)/U(1)$ group. The average values of the quasi-spin operators, expressed in terms of their stereographic projections \cite{per1, per2, omar}, are given by:
\begin{equation}
S^{+}=\bra{\psi}\hat{S}^{+}\ket{\psi}=\dfrac{\bar{\psi}}{1+|\psi| ^{2}}; \quad S^{-}=\bra{\psi}\hat{S}^{-}\ket{\psi}=\dfrac{{\psi}}{1+|\psi| ^{2}}; \quad S^{z}=\bra{\psi}\hat{S}^{z}\ket{\psi}=-\dfrac{1-|\psi| ^{2}}{2(1+|\psi| ^{2})}, \label{colm1}
\end{equation}
with $S ^{+}=S ^{x}+iS ^{y}$ and  $S ^{-}=S ^{x}-iS ^{y}$, where $\psi_{n}$, $\theta_{n}$ and $\phi_{n}$ are related by
\begin{equation}
\psi_{n}=\tan (\theta_{n}/2)e^{i \varphi_{n}}\label{dudie1},
\end{equation}
and their corresponding counterparts for the complementary strand. In addition, we consider the continuous limit 
\begin{equation}
\psi_{n+1}=\psi(z, t)\pm a\psi_{z}+\dfrac{a ^{2}}{2!}\psi_{zz}+..., \quad |\psi_{n\pm 1}| ^{2}=|\psi|\pm a\left(|\psi|\right)_{z}+\dfrac{a ^{2}}{2!}\left(|\psi|\right)_{zz}+..., \quad \sum_{n} \to \int \dfrac{dz}{a},
\end{equation}
similarly, for $X_{n+1}$, $Y_{n+1}$, $\xi_{n+1}$ and $|\xi_{n\pm 1}|$. This is possible once that the length of excitations in DNA is much greater than the inter-site distance $a$ between neighbouring nucleotides. Therefore, by averaging the Hamiltonian (\ref{darr4}) using the expressions in (\ref{colm1}) and incorporating the above considerations, we obtain:
\begin{eqnarray}
H &=& \int\Bigg[\frac{aJ}{2}\Bigg(\frac{|\psi_{z}|^{2}}{1+|\psi|^{2}}+\frac{|\xi_{z}|^{2}}{1+|\xi|^{2}}\Bigg)-\frac{1}{4a}(\mu-\alpha_{1}a(X_{z}))\frac{{2}(\bar{\psi}\xi+\bar{\xi}\psi)+(1-|\xi|^{2})(1-|\psi|^{2})}{(1-|\xi|^{2})(1-|\psi|^{2})}\nonumber\\
&& -\: \frac{\alpha_{2}}{2}\frac{(1-|\psi|^{2})(1-|\xi|^{2})}{(1+|\psi|^{2})(1+|\xi|^{2})}Y_{z}+\frac{p^{2}}{2am_{1}}+\frac{q^{2}}{2am_{2}}+k_{1}a(X_{z})^{2}+k_{2}a(Y_{z})^{2}\Bigg].\label{tay1}
\end{eqnarray}
Thus, the classical motion equations for the hydrogen atom displacement and the peptide displacement turns out to be
\begin{equation}
X_{z}=\dfrac{\alpha_{1}}{2(m_{1}v_{1}-2k_{1})}\dfrac{1+|\psi| ^{4}-6|\psi| ^{2}}{(1+|\psi| ^{2})^{2}} \label{jojo1}
\end{equation}
and
\begin{equation}
Y_{z}=\dfrac{\alpha_{2}}{2(m_{2}v_{2}-2k_{2})}\left(\dfrac{1-|\psi| ^{2}}{1+|\psi| ^{2}}\right) ^{2} \label{jojo2}
\end{equation}
with the governing equation for $\psi$ being
\begin{equation}
i \psi_{t}=-\psi_{ss}+\dfrac{2\mu}{a}\left(\dfrac{1-|\psi|^{2}}{1+|\psi|^{2}}\right)\psi-\left(\beta\dfrac{1+|\psi|^{4}-6|\psi|^{2}}{(1+|\psi|^{2})^{2}}+\gamma \left(\dfrac{1-|\psi|^{2}}{1+|\psi|^{2}}\right)^{2}\right)\dfrac{1-|\psi|^{2}}{1+|\psi|^{2}}\psi, \label{colm2}
\end{equation}
here, we consider the parametric change $z=s\sqrt{\frac{Ja}{2}}$, along with the condition $\frac{\mu}{a}\gg 1$. 

Additionally, we define the following parameters 
\begin{equation}
\beta=\frac{\alpha_{1} ^{2}}{2(m_{1}v_{1}^{2}-2k_{1})} \quad \text{and} \quad \gamma=\frac{\alpha_{2} ^{2}}{2(m_{2}v_{2}^{2}-2k_{2})},\label{adlle2}
\end{equation}
with $v_{1}$ and $v_{2}$ as velocities of the hydrogen atom and the peptide, respectively. In order to obtain (\ref{colm2}) we need to consider the symmetry of the base pairs in the complementary strands, i.e., $\psi=-\xi$ as well as Eqs. (\ref{jojo1}) and Eq. (\ref{jojo2}) (additional details can be found in \cite{max1}).  Equation (\ref{colm2}) represents a generalized NLSE with saturable nonlinearities, which fully describes the dynamics of the DNA-protein system.  The analytical solutions of the generalized NLSE in saturable media represent an intriguing challenge, and while some attempts have been made to solve the simplest versions of this equation directly \cite{saturable1, saturable2}, a general solution remains elusive. For the current study, however, we employ a weakly saturable approximation, where $G(I)=F(I)(1+I)^{-1}\approx F(I)(1-I)$ with $I=|\psi|^{2}$. After performing the reparametrization $\mu/a \to \mu$, we obtain the cubic-quintic NLSE, which is given by
\begin{equation}
i\psi_{t}+\psi_{ss}+k_{1}\psi+k_{3}|\psi|^{2}\psi+k_{5}|\psi|^{4}\psi=0, \label{frank1}
\end{equation}
with parameters 
\begin{equation}
k_{1}=-2\mu+\beta+\gamma, \quad k_{3}=4\mu-10\beta-6\gamma, \quad k_{5}=4\mu-34\beta-18\gamma. \label{adlle1}
\end{equation}

It is evident that the use of generalized coherent states associated with the $SU(2)/U(1)$ group provides a more general framework than that derived from the standard coherent states. This assertion can be readily demonstrated by showing that, with a careful choice of parameters, we can reduce our governing equation to previously studied cases, as we explore in the following limiting cases. 
\begin{enumerate}
\item By selecting the parameters $\mu$, $\beta$ and $\gamma$ such that $k_{5}=0$, we find that Eq. (\ref{frank1}) simplifies to the NLSE, which has been already obtained in the previously mentioned studies of open states of the base pairs \cite{vasum, danvas, gre1, dani, vasum2}. As it is well known the trivial boundary condition $x\to \pm \infty$ and $\psi \to 0$, with $k_{3}>0$ and normalized constants, lead to the a collective wave of bell type which form an open state with for the angle deviation, as indicated by Eq. (\ref{dudie1}).  This wave is expected to propagate along the DNA molecule due to interactions with thermal phonons surrounding the molecule. The collective wave is constructed from two types of bell-shaped solutions for the $\psi$ and $\xi$ fields, respectively.     

\item Another limiting case of Eq. (\ref{frank1}) arises when $k_{3}=0$. By selecting appropriate values of $\mu$, $\beta$, and $\gamma$ and $k_{5}>0$, we obtain the attractive Schrödinger equation with quintic nonlinearity or quintic Schrödinger equation. In this case, the evolution equation splits into two sectors. The first sector corresponds to collapse, where the amplitude of the solutions grows infinitely in a finite time. The second sector represents dispersion, where the amplitude decays and the localized region expands. The first sector holds no physical significance; therefore, effects that counteract this collapse must be introduced, such as inelastic collisions of DNA base pairs, which would lead to energy loss in the form of heat emission. 
\end{enumerate}

As shown earlier, our governing equation can indeed be directly reduced to the well-known cases. However, it also provides additional physical insights beyond what is already known. In addition to the classical soliton structures that form open states, as derived from the previously discussed cases, we also encounter non-classical soliton structures, such as compactons and anti-compactons. The compacton solution appears as a result of the interaction of two drop solitons for $\psi$ and $\xi$, once we replace the solutions in the expression for the hydrogen atom $X$ given by Eq. (\ref{jojo1}) with $\psi$ being the solution of Eq. (\ref{colm2}) with the boundary conditions $\psi\to 0$ as $x\to \pm \infty$. It's worth to point out that when the two solutions travel along the DNA chain the compacton forms open states and the anti-compacton breaks this deformation. Thus, the anticompactons play a major role in the DNA repair mechanism. 

Both classical and non-classical structures are highly sensitive to medium effects, which can substantially alter their profiles. As such, studying these structures is crucial for understanding the dynamics of open states in DNA chains, especially in relation to protein interactions.

\section{Perturbative analysis for the cubic-quintic NLSE solitons}

As it was already mentioned in the previous discussion, the DNA-protein system was treated as an isolated system supporting the cubic-quintic NLSE. However, in reality, the DNA molecule interacts with its surrounding environment. Modelling the interaction between a single DNA molecule and its environment is a complex problem, but in a simplified scenario, it can be reduced to two primary effects: dissipation and the influence of the endogenous field \cite{sat}. For this analysis, we assume that the solvent consists of water molecules, which dampen the vibrations of the DNA bases, leading to the dissipation of nonlinear excitations in the system. To model the damping effects, we apply the multi-scales perturbation approach proposed by Y. Kodoma and M. J. Ablowitz, commonly known as the quasi-stationary method \cite{ablow}. This method offers a significant advantage: it allows us to identify coherent, localized structures that emerge from the initial soliton-like solution. Before proceeding with the perturbative analysis of the molecular system, we first provide a brief overview of the quasi-stationary method.

\subsection{The quasi-stationary method}

Let us consider a perturbed nonlinear dispersive wave equation of the form
\begin{equation}
K(q, q_{t}, q_{x},...)=F(q, q_{x},...), \qquad 0<{\epsilon}\ll 1, \label{miles1}
\end{equation}
being $K$ and $F$ nonlinear functions of $q$, $q_{x},...$. We consider that the nonlinear equation (\ref{miles1}) depends implicitly of $\epsilon$, such that setting $\epsilon=0$ yields the unperturbed equation
\begin{equation}
K(q_{0}, q_{0t}, q_{0x},...) = 0,
\end{equation}
where $q_{0}$ is a solitary wave or a soliton solution. Now, we write the solution in terms of fast and slow variables:
\begin{equation}
q_{0}=\hat{q}_{0}(\theta_{1}, \theta_{2},..., \theta_{m}; P_{1}(T), P_{2}(T),..., P_{N}(T)). 
\end{equation} 
Here $\theta_{i} (i=1,...,m)$ are \textit{fast} variables, $T=\epsilon{t}$ represent \textit{slow} variables, and $P_{i}(i=1,...,N)$ are parameters that depend on the slow variables. Additionally, the variables  $\theta$  satisfy $\partial{\theta}/\partial{x}=1$, $\partial{\theta}/\partial{t}=-P_{1}$ and $P_{1}=P_{1}(T)$ in order to remove secular terms. Therefore, these solutions are called quasi-stationary solutions $q=\hat{q}(\theta, T, \epsilon)$. Thus, we assume an expression of the form 

\begin{equation}
\hat{q}=\hat{q}_{0}+{\epsilon}\hat{q}_{1}+{\epsilon}^{2}\hat{q}_{2}+...
\end{equation} 
Substituting $\hat{q}$ into Eq. ($\ref{miles1}$) yields the first-order equations:

\begin{equation}
L(\partial_{\theta_{i}}, \hat{q}_{0})\hat{q}_{1}{\equiv}F(\hat{q}_{0})-\frac{\partial{K}}{\partial{q_{t}}}\cdot{q_{T}}|_{q=\hat{q}_{0}}{\equiv}\hat{F},
\end{equation}
where $L(\partial_{\theta_{i}}, \hat{q}_{0})u=0$ is a linearized equation of $K(q, q_{t}, q_{x},...)=0$. We can denote by $v_{i}(i=1,...,M)$ to the $M$ solutions of the homogeneous adjoint problem satisfying 

\begin{equation}
L^{A}v_{i}=0, \qquad i=1,..., M, \qquad M{\leq}N,
\end{equation}
with $L^{A}$ as the adjoint operator of $L$, we have

\begin{equation}
({L}\hat{q}_{i})\cdot{v_{i}}-(L^{A}v_{i})\cdot{\hat{q}_{1}}=\hat{F}v_{i}.
\end{equation} 
If the last equation can be integrated to yield the secularity conditions, we will be in a position to compute the solution for $\hat{q}_{1}$. 

\subsection{First order perturbation for the  drop-soliton solution of  the CQNSE}

To implement the quasi-stationary method outlined above to our DNA-protein system, we begin with the perturbed form of the cubic-quintic NLSE
\begin{equation}
i\varphi_{t}+\frac{1}{2}\varphi_{xx}+{\varphi}{|\varphi|^{2}}-\alpha\varphi{|\varphi|^{4}}=-i{\epsilon}\varphi,  \label{psc}
\end{equation}
which was obtained by the following scale transformation of Eq. (\ref{frank1})
\begin{equation}
\psi=e^{ik_{1}t}\phi(s, t), \  \phi=\varphi (\beta s, \beta^2 {t}), \label{m1}
\end{equation}
with $\beta=(k_{3}/2)^{1/2}$ and $\alpha=-2k_{5}/k_{3}$, we also introduce the reparametrization $s \to x$.

The right side of the equation (\ref{psc}) describes the damping effect caused by the viscous medium. The parameter   $0<\epsilon\ll 1$ is a coupling coefficient of the strength of viscous damping, which is very small at physiological temperatures. Considering the previous assumption, and in order to investigate the evolution of nonlinear excitations in the DNA molecular chain, we employ the perturbative analysis outlined in the preceding section. When $\epsilon=0$, we recover the unperturbed cubic-quintic NLSE, which admits the following drop soliton-like solution \cite{serk1}
\begin{equation}
\varphi=\frac{A\exp(iV(\theta-\theta_{0})+i(\sigma-\sigma_{0})}{\bigg[1+\sqrt{1-\frac{4}{3}\alpha{A^{2}}}\, \text{cosh}(\sqrt{2}A(\theta-\theta_{0})\bigg]^{1/2}}, \label{lau1}
\end{equation}
where
\begin{equation}
\frac{\partial{\sigma}}{\partial{t}}=\frac{A^{2}}{4}+\frac{V^{2}}{2}, \qquad \frac{\partial{\sigma}}{\partial{x}}=0, \qquad \frac{\partial{\theta}}{\partial{x}}=1, \qquad \frac{\partial{\theta}}{\partial{t}}=-V.
\end{equation}
Using the quasi-stationary method, we express the soliton-like solution (\ref{lau1}) in terms of both fast and slow variables. To this end, we introduce a slow time variable $T={\epsilon}t$, and  $A$, $V$, $\theta_{0}$ ,  $\sigma_{0}$ being functions of this new time scale. This allows us to express the envelope of the one-soliton solution as follows:
\begin{equation}
\varphi=\hat{\varphi}(\theta, T; \epsilon)exp[iV(\theta-\theta_{0})+i(\sigma-\sigma_{0})]. \label{ge}
\end{equation}
Substituting (\ref{ge}) into (\ref{psc}) and assuming quasi-stationarity, we obtain:
\begin{equation}
-\frac{A^{2}}{4}\hat{\varphi}+\frac{1}{2}\hat{\varphi}_{{\theta}\theta}+|\hat{\varphi}|^{2}\hat{\varphi}-{\alpha}|\hat{\varphi}|^{4}\hat{\varphi}={\epsilon}F(\hat{\varphi}), \label{a1}
\end{equation} 
where

\begin{equation}
F(\hat{\varphi})=-i(\hat{\varphi}+\hat{\varphi}_{T})+[(\theta-\theta_{0})V_{T}-V\theta_{0T}-\sigma_{0T}]\hat{\varphi}. \label{a2}
\end{equation}
We assume that $\hat{\varphi}$ is expressed as
\begin{equation}
\varphi=\hat{\varphi}_{0}+\epsilon\hat{\varphi}_{1}+{\epsilon}^{2}\hat{\varphi}_{2}+...
\end{equation} 
Neglecting higher order of $\epsilon$, we focus on the first-order perturbation, such that
\begin{equation}
\varphi=\hat{\varphi}_{0}(\theta, T)+\epsilon\hat{\varphi}_{1}, \label{gol1}
\end{equation}
where
\begin{equation}
\hat{\varphi}_{0}(\theta, T)=\frac{A}{\bigg[1+\sqrt{1-\frac{4}{3}\alpha{A^{2}}}\, \text{cosh}(\sqrt{2}A(\theta-\theta_{0})\bigg]^{1/2}} \label{gol2}
\end{equation}
and $\hat{\varphi}_{1}=\phi_{1}+{i}\psi$, being $\phi_{1}$ and $\psi_{1}$ real functions. Substituting expressions (\ref{gol1}) and (\ref{gol2}) into Eqs. (\ref{a1}) and (\ref{a2}), we obtain the following system of equations
\begin{subequations}
\begin{align}
L_{1}\phi_{1} & {\equiv}\frac{1}{2}\phi_{{1}\theta\theta}+3|\varphi_{0}|^{2}\phi_{1}-5{\alpha}|\varphi_{0}|^{4}\phi_{1}-\frac{A^{2}}{4}\phi_{1}=\text{Re}F(\hat{\varphi}_{0}), \label{d1}\\
L_{2}\psi_{1} & {\equiv}\frac{1}{2}\psi_{{1}\theta\theta}+|\varphi_{0}|^{2}\psi_{1}-{\alpha}|\varphi_{0}|^{4}\psi_{1}-\frac{A^{2}}{4}\psi_{1}=\text{Im}F(\hat{\varphi}_{0}),\label{d2}
\end{align}
\end{subequations}
with 
\begin{subequations}
\begin{align}
\text{Re}F(\hat{\varphi}_{0}) &= [(\theta-\theta_{0})V_{T}-V\theta_{0T}-\sigma_{0T}]\hat{\varphi}_{0},\label{d3}\\
\text{Im}F(\hat{\varphi}_{0}) &= -[\hat{\varphi}_{0T}+\hat{\varphi}_{0}].\label{d4}
\end{align}
\end{subequations}
The operators $L_{1}$ and $L_{2}$ are self-adjoint and $L_{1}\hat{\varphi}_{0\theta}=L_{2}\hat{\varphi}_{0}=0$. Consequently,  the equations (\ref{d1}) and (\ref{d2}) have localized solutions around the drop soliton. The conditions of solvability of the system are given by the following secularity conditions
\begin{subequations}
\begin{align}
\int_{-\infty}^{\infty}\hat{\varphi}_{0{\theta}}\text{Re}(\hat{\varphi}_{0})d\theta=0,\\
\int_{-\infty}^{\infty}\hat{\varphi}_{0}\text{Im}(\hat{\varphi}_{0})d\theta=0,
\end{align}
\end{subequations}
upon integrating the above expressions, we obtain
\begin{equation}
\frac{\partial{V}}{\partial{T}}=0, \qquad \mbox{ and }\qquad \frac{\partial{A}}{\partial{T}}=\frac{\sqrt{2}\epsilon(3-4{\alpha}{A}^{2}){E}}{3}, \label{delia1}
\end{equation}
where $E=\int_{-\infty}^{\infty}|\varphi|^{2}dx$ is the energy of the soliton. 

\medskip

\begin{figure}[H]
\includegraphics[width=1.0\linewidth]{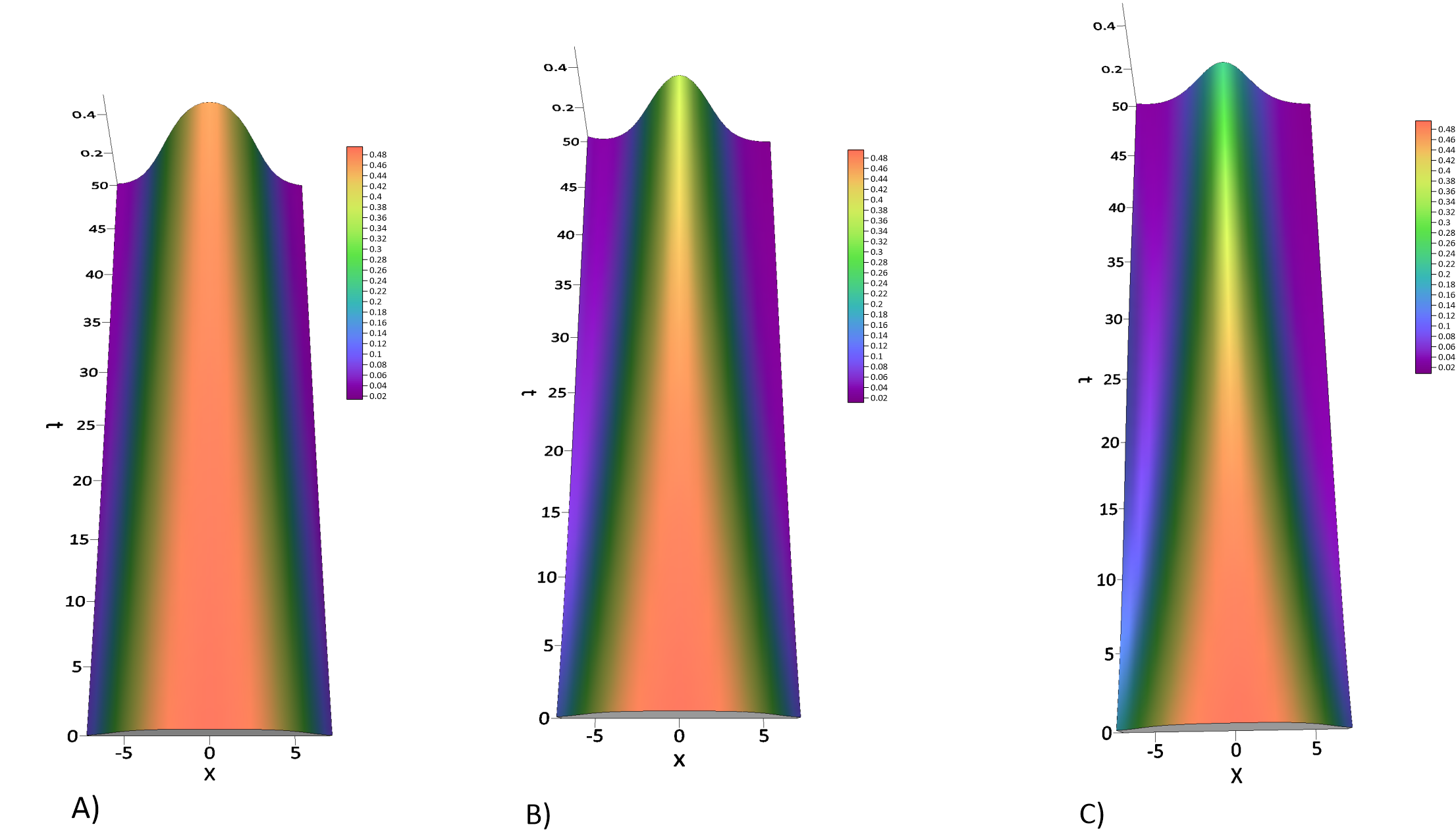}
\centering
\caption{Quasi-stationary evolution of the soliton-like solution $|{\varphi}|^{2}$ for a) $\epsilon=0.01$, b) $\epsilon=0.02$ and c) $\epsilon=0.03$, respectively. }
\label{fig:tak4}
\end{figure}

In the presence of the perturbation term the parameters deform adiabatically, these adiabatic variations are given by (\ref{delia1}). This is an easy way to corroborate the equations obtained by the Fredholm alternative.    
In order to find the perturbed soliton solutions, we need to solve the homogeneous part of (\ref{d1}) and (\ref{d2}). Starting from (\ref{d1}),  we can easily see that the solutions are:
\begin{equation}
\phi_{11} = \frac{-A^{2}\sqrt{1-\frac{4}{3}\alpha{A^{2}}}\, \text{sinh}\left(\sqrt{2}A(\theta-\theta_{0})\right)}{\sqrt{2}\bigg[1+\sqrt{1-\frac{4}{3}\alpha{A^{2}}}\, \text{cosh}\left(\sqrt{2}A(\theta-\theta_{0})\right)\bigg]^{3/2}},\label{r1}
\end{equation}
\begin{eqnarray}
\phi_{12} &= A\sqrt{1-\frac{4}{3}\alpha{A^{2}}}\Bigg\{\frac{(1-2\alpha{A^{2}})\, \text{cosh}\left(\sqrt{2}A(\theta-\theta_{0})\right)-2(1-\frac{1}{3})\alpha{A^{2}}\sqrt{1-\frac{4}{3}\alpha{A^{2}}}}{\big[1+\sqrt{1-\frac{4}{3}\alpha{A^{2}}}\, \text{cosh}(\sqrt{2}A(\theta-\theta_{0})\big]^{3/2}}\nonumber\\
& -\: \frac{A}{\sqrt{2}}(1-\frac{4}{3}\alpha{A^{2}})\frac{(\theta-\theta_{0})\, \text{sinh}\left(\sqrt{2}A(\theta-\theta_{0})\right)}{\big[1+\sqrt{1-\frac{4}{3}\alpha{A^{2}}}\, \text{cosh}\left(\sqrt{2}A(\theta-\theta_{0})\right)\big]^{3/2}} \Bigg\}. \label{r2}
\end{eqnarray}
Using Eqs. (\ref{r1}) and (\ref{r2}), together with the secularity condition Eq. (\ref{delia1}), we obtain the following solution
\begin{eqnarray}
\phi_{1} =& C_{1}\phi_{11}+C_{2} \phi_{12}+N_{1}\Bigg\{\frac{A^{4}(1-2{\alpha}A^{2})(1-\frac{4}{3}\alpha{A^{2}})}{\frac{8}{3}\alpha{A^{2}}}-\frac{2A^{4}(1-\frac{1}{3}\alpha{A^{2}})(1-\frac{4}{3}\alpha{A^{2}})^{2}}{\frac{8}{3}\alpha{A^{2}}}\Bigg\}(V\theta_{0T}+\sigma_{0T})\nonumber\\
& +\: (N_{3}-N_{2})\Bigg\{\frac{A^{4}(1-2{\alpha}A^{2})(1-\frac{4}{3}\alpha{A^{2}})^{3/2}}{(\sqrt{2})^{2}(\frac{4}{3}\alpha{A^{2}})^{3/2}}-\frac{2A^{4}(1-\frac{1}{3}\alpha{A^{2}})(1-\frac{4}{3}\alpha{A^{2}})^{3/2}}{(\sqrt{2})^{2}\frac{4}{3}\alpha{A^{2}})^{3/2}}+\frac{\frac{A^{5}}{2}(1-\frac{4}{3}\alpha{A^{2}})^{3/2}}{\sqrt{2}(\sqrt{2}A)^{2}\sqrt{\frac{4}{3}\alpha{A^{2}}}}\Bigg\}(V{\theta}_{0T}+\sigma_{0T})\nonumber\\
& +\: N_{4}\Bigg[\frac{\frac{A^{6}}{2}(1-\frac{4}{3}\alpha{A^{2}})^{5/2}}{\sqrt{2}(\sqrt{2}A)^{2}}\Bigg](V\theta_{0T}+\sigma_{0T}) \label{dulce1}
\end{eqnarray}
with $\frac{4}{3}\alpha{A^{2}}>0$, where $C_{1}$ and $C_{2}$ are arbitrary constants. Additionally, $N_{1}$, $N_{2}$, $N_{3}$ and $N_{4}$ are defined as follows
$$N_{1}=\frac{\text{sinh}^{2}\sqrt{2}\left(A(\theta-\theta_{0})\right)}{\sqrt{2}\bigg[1+\sqrt{1-\frac{4}{3}\alpha{A^{2}}}\, \text{cosh}\left(\sqrt{2}A(\theta-\theta_{0})\right)\bigg]^{5/2}},$$
$$N_{2}=\frac{\text{sinh}\left(\sqrt{2}A(\theta-\theta_{0})\right)}{\bigg[1+\sqrt{1-\frac{4}{3}\alpha{A^{2}}}\, \text{cosh}\left(\sqrt{2}A(\theta-\theta_{0})\right)\bigg]^{3/2}}\ln\left[\frac{1+\sqrt{1-\frac{4}{3}\alpha{A^{2}}}+\sqrt{\frac{4}{3}\alpha{A^{2}}}\, \text{tanh}\left(\frac{\sqrt{2}A(\theta-\theta_{0})}{2}\right)}{1+\sqrt{1-\frac{4}{3}\alpha{A^{2}}}-\sqrt{\frac{4}{3}\alpha{A^{2}}}\, \text{tanh}\left(\frac{\sqrt{2}A(\theta-\theta_{0})}{2}\right)}\right],$$
$$N_{3}= \frac{\text{sinh}\left(\sqrt{2}A(\theta-\theta_{0})\right)}{\bigg[1+\sqrt{1-\frac{4}{3}\alpha{A^{2}}}\, \text{cosh}\left(\sqrt{2}A(\theta-\theta_{0})\right)\bigg]^{3/2}}\ln\left[\frac{1+\sqrt{1-\frac{4}{3}\alpha{A^{2}}}-\sqrt{\frac{4}{3}\alpha{A^{2}}}}{1+\sqrt{1-\frac{4}{3}\alpha{A^{2}}}+\sqrt{\frac{4}{3}\alpha{A^{2}}}}\right],$$
$$N_{4}=\frac{\text{sinh}\left(\sqrt{2}A(\theta-\theta_{0})\right)}{\bigg[1+\sqrt{1-\frac{4}{3}\alpha{A^{2}}}\, \text{cosh}\left(\sqrt{2}A(\theta-\theta_{0})\right)\bigg]^{5/2}}.$$
Now, we solve the homogeneous part of equation (\ref{d2}) obtaining
\begin{subequations}
\begin{align}
\psi_{11}=\frac{A}{\big[1+\sqrt{1-\frac{4}{3}\alpha{A^{2}}}\, \text{cosh}\left(\sqrt{2}A(\theta-\theta_{0})\right)\big]^{1/2}} \label{dia1},   \\                               
\psi_{12}=\frac{\sqrt{2}A(\theta-\theta_{0})+\sqrt{1-\frac{4}{3}\alpha{A^{2}}}\, \text{sinh}\left(\sqrt{2}A(\theta-\theta_{0})\right)}{\sqrt{2}\big[1+\sqrt{1-\frac{4}{3}\alpha{A^{2}}}\, \text{cosh}\left(\sqrt{2}A(\theta-\theta_{0})\right)\big]^{1/2}}. \label{dia2}
\end{align}
\end{subequations}
We can derive the general solution as before by applying the method of variation of parameters, which leads to Eqs. (\ref{dia1}) and (\ref{dia2}). Consequently, the general solution for $\psi_{1}$ is found to be: 
\begin{eqnarray}
\psi_{1} &= C_{3}\psi_{11}+C_{4}\psi_{12}+\Bigg[\frac{A^{2}}{\sqrt{2}\sqrt{\frac{4}{3}\alpha{A^{2}}}}\Bigg](M_{1}-M_{2})-\frac{A^{2}\theta_{0T}}{2\sqrt{2}}M_{3}+\Bigg[\frac{A}{2\sqrt{2}}\Bigg]M_{4}- A^{5}\theta_{0T}M_{5}\nonumber\\
& +\: \frac{(1-\frac{4}{3}\alpha{A^{2}}){A^{3}}\theta_{0T}}{{2}(\frac{4}{3}\alpha{A^{2}})}M_{6}+\Bigg[\frac{A^{4}\theta_{0T}}{\sqrt{2}\sqrt{\frac{4}{3}\alpha}}+\frac{A}{(\frac{4}{3}\alpha{A^{2}})^{3/2}}\Bigg](M_{7}-M_{8})\label{dulcef2}
\end{eqnarray}
with $\frac{4}{3}\alpha{A^{2}}>0$. The expressions for $M({n})$, where $n{\leq}8$, are given by

$$M_{1}=\frac{\sqrt{2}A(\theta-\theta_{0})+\sqrt{1-\frac{4}{3}\alpha{A^{2}}}\, \text{sinh}\left(\sqrt{2}A(\theta-\theta_{0})\right)}{\sqrt{2}\big[1+\sqrt{1-\frac{4}{3}\alpha{A^{2}}}\, \text{cosh}\left(\sqrt{2}A(\theta-\theta_{0})\right)\big]^{1/2}}\ln\left[\frac{1+\sqrt{1-\frac{4}{3}\alpha{A^{2}}}+\sqrt{\frac{4}{3}\alpha{A^{2}}}\,\text{tanh}\left(\frac{\sqrt{2}A(\theta-\theta_{0})}{2}\right)}{1+\sqrt{1-\frac{4}{3}\alpha{A^{2}}}-\sqrt{\frac{4}{3}\alpha{A^{2}}}\,\text{tanh}\left(\frac{\sqrt{2}A(\theta-\theta_{0})}{2}\right)}\right],$$

$$M_{2}=\frac{\sqrt{2}A(\theta-\theta_{0})+\sqrt{1-\frac{4}{3}\alpha{A^{2}}}\, \text{sinh}\left(\sqrt{2}A(\theta-\theta_{0})\right)}{\sqrt{2}\big[1+\sqrt{1-\frac{4}{3}\alpha{A^{2}}}\, \text{cosh}\left(\sqrt{2}A(\theta-\theta_{0})\right)\big]^{1/2}}\ln\left[\frac{1+\sqrt{1-\frac{4}{3}\alpha{A^{2}}}-\sqrt{\frac{4}{3}\alpha{A^{2}}}}{1+\sqrt{1-\frac{4}{3}\alpha{A^{2}}}+\sqrt{\frac{4}{3}\alpha{A^{2}}}}\right],$$

$$M_{3}=\frac{\sqrt{2}A(\theta-\theta_{0})+\sqrt{1-\frac{4}{3}\alpha{A^{2}}}\, \text{sinh}\left(\sqrt{2}A(\theta-\theta_{0})\right)}{\big[1+\sqrt{1-\frac{4}{3}\alpha{A^{2}}}\, \text{cosh}\left(\sqrt{2}A(\theta-\theta_{0})\right)\big]^{3/2}}, \quad M_{4}=\frac{\ln\left(1+\sqrt{1-\frac{4}{3}\alpha{A^{2}}}\, \text{cosh}\left(\sqrt{2}A(\theta-\theta_{0})\right)\right)}{\big[1+\sqrt{1-\frac{4}{3}\alpha{A^{2}}}\, \text{cosh}\left(\sqrt{2}A(\theta-\theta_{0})\right)\big]^{1/2}}, $$

$$M_{5}=\frac{\theta-\theta_{0}}{\big[1+\sqrt{1-\frac{4}{3}\alpha{A^{2}}}\, \text{cosh}\left(\sqrt{2}A(\theta-\theta_{0})\right)\big]^{3/2}}, \quad M_{6}=\frac{\text{sinh}\left(\sqrt{2}A(\theta-\theta_{0})\right)}{\big[1+\sqrt{1-\frac{4}{3}\alpha{A^{2}}}\, \text{cosh}\left(\sqrt{2}A(\theta-\theta_{0})\right)\big]^{3/2}},$$

$$M_{7}=\frac{1}{\big[1+\sqrt{1-\frac{4}{3}\alpha{A^{2}}}\, \text{cosh}\left(\sqrt{2}A(\theta-\theta_{0})\right)\big]^{1/2}} \ln\left[\frac{1+\sqrt{1-\frac{4}{3}\alpha{A^{2}}}+\sqrt{\frac{4}{3}\alpha{A^{2}}}\,\text{tanh}\left(\frac{\sqrt{2}A(\theta-\theta_{0})}{2}\right)}{1+\sqrt{1-\frac{4}{3}\alpha{A^{2}}}-\sqrt{\frac{4}{3}\alpha{A^{2}}}\, \text{tanh}\left(\frac{\sqrt{2}A(\theta-\theta_{0})}{2}\right)}\right],$$

$$M_{8}=\frac{1}{\big[1+\sqrt{1-\frac{4}{3}\alpha{A^{2}}}\, \text{cosh}\left(\sqrt{2}A(\theta-\theta_{0})\right)\big]^{1/2}} \ln\left[\frac{1+\sqrt{1-\frac{4}{3}\alpha{A^{2}}}-\sqrt{\frac{4}{3}\alpha{A^{2}}}}{1+\sqrt{1-\frac{4}{3}\alpha{A^{2}}}+\sqrt{\frac{4}{3}\alpha{A^{2}}}}\right].$$ 
Now, let us consider $\theta_{0}=\sigma_{0}=0$ in Eq. (\ref{lau1}), which correspond to the center of the soliton and the soliton phase, respectively. Additionally, from the secularity conditions Eq. (\ref{delia1}), we have $V_{T}=0$. Thus, for Eqs. (\ref{dulce1}) and (\ref{dulcef2}), we obtain:
\begin{equation}
\phi_{1}=C_{1}\phi_{11}+C_{2}\phi_{12}. \label{lau3}
\end{equation}
Imposing the boundary conditions $\phi_{1}|_{\theta=0}=C$ and $\phi_{{1}\theta}|_{\theta=0}=0$, we find the values for the arbitrary constants to be

$$C_{1}=0, \quad \text{and} \quad C_{2}=\frac{C\Big[1+\sqrt{1-\frac{4}{3}\alpha{A^{2}}}\Big]^{3/2}}{A\sqrt{1-\frac{4}{3}\alpha{A^{2}}}\{(1-2\alpha{A^{2}})-2(1-\frac{1}{3})\alpha{A^{2}}\sqrt{1-\frac{4}{3}\alpha{A^{2}}}\}}.$$
Similarly, for Eq. (\ref{lau3}), we obtain that
\begin{equation}
\psi_{1}=C_{3}\psi_{11}+C_{4}\psi_{12}+\Bigg[\frac{A^{2}}{\sqrt{2}\sqrt{\frac{4}{3}\alpha{A^{2}}}}\Bigg](M_{1}-M_{2})+\Bigg[\frac{A}{2\sqrt{2}}\Bigg]M_{4}+\frac{A}{(\frac{4}{3}\alpha{A^{2}})^{3/2}}(M_{7}-M_{8}). \label{lau4}
\end{equation}
Imposing the boundary conditions $\psi_{1}|_{\theta=0}=0$ and $\psi_{{1}\theta}|_{\theta=0}=0$, we obtain  

$$C_{3}= \frac{1}{\bigg(\frac{4}{3}\alpha{A^{2}}\bigg)^{3/2}}\ln\left[\frac{1+\sqrt{1-\frac{4}{3}\alpha{A^{2}}}-\sqrt{\frac{4}{3}\alpha{A^{2}}}}{1+\sqrt{1-\frac{4}{3}\alpha{A^{2}}}+\sqrt{\frac{4}{3}\alpha{A^{2}}}} \right]-\frac{1}{2\sqrt{2}}\ln\bigg[1+\sqrt{1-\frac{4}{3}\alpha{A^{2}}}\bigg]$$
and

$$C_{4}=
\frac{A^{2}}{\sqrt{2}\sqrt{\frac{4}{3}}\alpha{A^{2}}} \ln\left[\frac{1+\sqrt{1-\frac{4}{3}\alpha{A^{2}}}-\sqrt{\frac{4}{3}\alpha{A^{2}}}}{1+\sqrt{1-\frac{4}{3}\alpha{A^{2}}}+\sqrt{\frac{4}{3}\alpha{A^{2}}}}\right].$$
Finally, we can see that the general solution of the perturbed cubic-quintic NLSE is given by
\begin{equation}
\varphi_{1}=[\varphi_{0}+\epsilon(\phi_{1}+{i}\psi_{1})]\exp(iV(\theta-\theta_{0})+i(\sigma-\sigma_{0})), \label{o1}
\end{equation}
with $\phi_{1}$ and $\psi_{1}$ given by Eq. (\ref{dulce1}) and Eq. (\ref{dulcef2}), correspondingly. From the solution in Eq. (\ref{o1}), it is evident that under perturbative effects, the drop soliton solution of Eq. (\ref{psc}) decays over time, as shown in Fig. \ref{fig:tak4}. We observe that the amplitude of the soliton decreases as time progresses, a consequence of the secularity conditions, while its velocity remains constant. Additionally, from the figures, we can note that the damping effect increases over a short period when the damping coefficient is larger. In general, the perturbed term in Eq. (\ref{psc}) can take the form of any function $R(\varphi)$. For the specific case of DNA internal dynamics, we use $R(\varphi)=\varphi$ to model the effect of a viscous medium.  A similar approach is seen in the work by A. Biswas, where a different form of $R(\varphi)$ is considered for unchirped fibres in the context of the cubic-quintic NLSE \cite{bis}.

We can also note from our analysis that the higher-order nonlinearity lead to unusual and contrasting effects of the soliton self-spreading (or self-compression), due to their amplification (or absorption). The soliton can be compressed as it travels in an absorbing medium and it can self-spread in its adiabatic amplification. This effect, which was original studied in the context of optical fibres \cite{serk0}, cannot be accomplish with the standard coherent states approach once it leads to a NLSE. 

To assess the impact of the viscous medium on the displacement of hydrogen bonds and peptides, we need to substitute Eq. (\ref{o1}) into Eq. (\ref{m1}), using the explicit forms given by Eqs. (\ref{jojo1}) and (\ref{jojo2}), respectively. It is expected that the quasi-stationary evolution of drop solutions, as shown in Fig. \ref{fig:tak4}, will directly affect non-classical solitonic structures, such as compactons and anti-compactons. This will result in deformations of their profiles, leading to either a fast or slow decrease in amplitude, as well as compression of the soliton's width. The latter will depend on whether the term corresponds to a loss or gain, as indicated by the sign of the viscous term. Another important, though less direct, consequence is the potential emergence of new soliton structures arising from the perturbed solution. As demonstrated in previous studies, the implementation of the quasi-stationary approach directly impacts the formation of novel structures in solitons embedded in lipid membranes \cite{omm1}. However, the complex mathematical forms of the perturbed soliton in the DNA-protein system will require a more detailed analysis to confirm our main hypothesis. To ensure the robustness of our findings, this analysis will be addressed in a future work.

\section{Some comments on the process of DNA transcription}
The analytical results discussed above can be applied to the transcription process, which is widely recognized for producing mRNA that carries the genetic information required for protein synthesis. In addition to mRNA, transcription also produces other essential RNA molecules, including transfer RNA (tRNA), ribosomal RNA (rRNA), and various non-coding RNAs, each of which plays a vital role in cellular structure and catalysis. All of these RNA molecules are synthesized by RNA polymerase enzymes, which transcribe DNA sequences into RNA copies. The enzyme binds to the promoter region of the DNA and initiates transcription at a specific start site within the promoter. Then it synthesizes the RNA strand, progressing along the DNA template until it encounters a termination signal. Upon reaching this signal, both the RNA polymerase and the newly synthesized RNA molecule are released. Therefore, the synthesis of an RNA chain consisting of 5,000 nucleotides takes approximately 3 minutes to complete \cite{alberts}. The DNA binds to the active site of the protein, causing the base pairs to separate and unwinding the DNA double helix.
Recently, numerical studies have been made to understand the RNAp-DNA  dynamics taking into account the helical coupling and the inhomogeneties of DNA \cite{kofa1, kofa2}. 
\begin{figure}[H]
\includegraphics[width=0.6\linewidth]{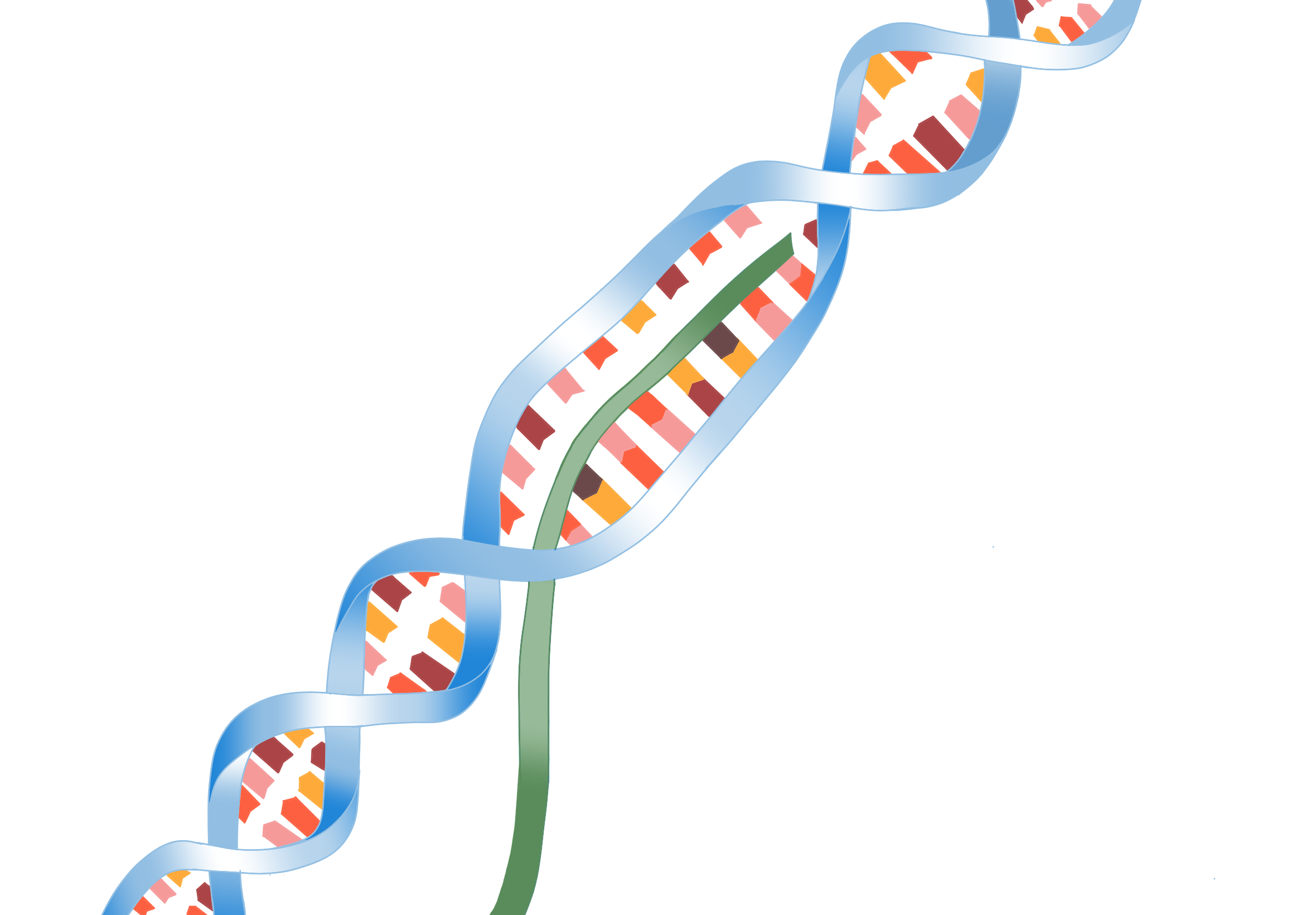}
\centering
\caption{Schematic representation of the transcription process. The transcription process begins with the double-helix unwinding and unzipping (as in the replication process).}
\label{fig:tak2}
\end{figure}
In this study, we neglect contributions of inhomogeneities once that the effect of these contributions is small compared to the homogeneous stacking. Besides, in these studies the localized inhomogeneity does not affect the bubble soliton sizes, even when the amplitude of inhomogeneity increases and the number of base pairs participating in the opening does not change due to this pattern of inhomogeneities. Moreover, we are interested in the analytic evolution of soliton-like solutions in the DNA-protein system under the viscous medium.
From these numerical studies \cite{kofa1, kofa2} we obtained the approximate values of the hydrogen bonds coefficient ($\mu$) and the stacking energy costant($J$), the elastic constants ($k_{1}$) of the RNAp molecule and phonon chain ($k_{2}$) and their coupling coefficients to the DNA molecule ($\alpha_{1}$ and $\alpha_{2}$), respectively. Also, the mass of the hydrogen atom attached to the base ($m_{1}$) and the mass of the peptide ($m_{2}$) are well-known.

$$J=1.5 \ \ eV \ \ \si{\angstrom}^{-2}; \ \ \ \ k_{1}=1.5 \ \ eV \ \ \si{\angstrom}^{-2};$$

$$\mu=0.1 \ \ eV; \ \ \ \ \alpha_{1}=0.005 \ \ eV; $$

$$\alpha_{2}=0.081 \ \ eV; \ \ \ \ m_{1}=1.0079 \ \ a. m. u;$$

$$m_{2}=300 \ \ a. m. u.; \ \ \ \ a= 3.4 \ \ \si{\angstrom}  $$
Besides, it is known that the cubic-quintic NLSE has several \textit{vacuum states}, i.e, classical configurations with minimal energy. Therefore, the solution of the equation of motion that lead to asymptotic values of $\psi(s)$ in Eq. (\ref{frank1}) has to coincide with minimum potentials \cite{max1}.  Thus, we find that the velocities of the displacement of the peptide and the hydrogen atom are given by:
\begin{equation}
\dfrac{A}{\rho_{0}}=-2-\dfrac{3}{4}\dfrac{k_{3}^{2}}{k_{1}k_{5}}\Bigg(1 \pm \sqrt{1+4\dfrac{k_{1}k_{5}}{k_{3}^{2}}}\Bigg)\label{joy},
\end{equation}
without any loss of generality we fix $\rho_{0}=1$, once that the solution properties can be represented as combinations of $A/\rho_{0}$ with $A>1$. Consequently, we can choose arbitrary $v_{1}$ and $v_{2}$ that satisfies the condition given by Eq. (\ref{joy}). 

In previous studies, S. Sdravkovi\'c and M. V. Satari\'c \cite{sat} determined, using the Peyrard's DNA model, that the speed and width of the soliton-like solution are $3750 \ m/s$ and $580 \si{\angstrom}$, respectively. This implies that the soliton spans approximately 170 bases in each DNA strand. The number of base pairs involved in the opening process depends on the nature of the soliton; however, it is preferable for fewer base pairs to participate in the opening process.
Finally, the value of the coupling coefficient for the strength of viscous damping, $\epsilon$,  is temperature-dependent, typically around physiological temperatures ($\sim 300$ K). Therefore, $\epsilon$ is of the order of $10^{-11} kg/s$, with the experimental value being approximately $\epsilon\approx 6 \times 10^{-11} kg/s$.

\section{Conclusion}

In this study, we analyzed the nonlinear dynamics of DNA and the evolution of nonlinear excitations induced by protein interactions and the surrounding viscous medium, using a quasi-stationary approach. Unlike previous studies that modeled DNA as a rigid molecular chain governed by the NLSE, we employed a cubic-quintic NLSE to account for the flexibility of DNA strands and the physiological properties of the protein-DNA binding system. We did not consider helicoidal terms in our model, as numerical simulations using the Yakushevich model have shown that their impact on soliton-like solutions is minimal. The viscous medium introduces damping effects on the nonlinear excitations, reducing the amplitude and resulting in a constant velocity for the drop-like soliton. As the damping coefficient $\epsilon$ increases, this damping effect becomes more pronounced, causing the soliton to decay more rapidly over time. For pure internal DNA dynamics, where $\alpha_{1}=\alpha_{2}=0$, we introduce various perturbations, such as periodic or pulse-like disturbances, which arise due to localized inhomogeneities in the DNA strands. While this work focuses on the linear form of $R(\varphi)$, similar analyses can be applied to other localized perturbations, both in DNA and in any system governed by the cubic-quintic NLSE. One of the key advantages of the quasi-stationary approach is its versatility in analyzing various soliton-like solutions, such as kinks, breathers, drops, and bubbles. Although this approach provides a useful approximation for understanding the behavior of DNA energy and information encoded in the strands, a more in-depth discrete analysis is necessary to fully capture the complexities of the double-stranded DNA system. Additionally, the direct impact of quasi-stationary evolution on compacton and anti-compacton solitons, which describe the open states in DNA chains, remains underexplored, presenting an opportunity for future research. Nevertheless, the present study provides explicit solutions for the direct analysis of these deformations.
\section*{Acknowledgments}
OPT acknowledges CONAHCyT by a postdoctoral fellowship.

\medskip

\noindent \textbf{Data Availability Statement} Data sharing not applicable to this article as no datasets were generated or analyzed during the current study.

\end{document}